\documentclass[12pt]{article}
\usepackage[esperanto]{babel}
\usepackage{epsfig} 

\newcommand{\bea}{\begin{eqnarray}}
\newcommand{\eea}{\end{eqnarray}}
\newcommand{\dd}{\mathrm{d}}

\begin{document}
\selectlanguage{esperanto} 

\title{\bf Relativeca Dopplera efiko inter du akcelataj korpoj -- I}  
\author{F.M. Paiva \\ 
{\small Departamento de F\'\i sica, U.E. Humait\'a II, Col\'egio Pedro II} \\
{\small Rua Humait\'a 80, 22261-040  Rio de Janeiro-RJ, Brasil; fmpaiva@cbpf.br} 
\vspace{.7ex} \\
A.F.F. Teixeira \\
{\small Centro Brasileiro de Pesquisas F\'\i sicas} \\
{\small 22290-180 Rio de Janeiro-RJ, Brasil; teixeira@cbpf.br}} 

\maketitle 

\begin{abstract} 
Ni priskribas luman Doppleran efikon inter same akcelataj fonto kaj observanto ^ce special-relativeco. La propraj akceloj estas konstantaj kaj paralelaj.  \\ - - - - - - - - - - - \\ 
We describe the Doppler effect between equally accelerated light source and observer under the special relativity. The proper accelerations are constant and parallel. An English version is available by request.
\end{abstract}

 
\section{Enkonduko}                                             \label{e}

Lorentzaj transformoj estas gravaj en special-relativeco. Ili montras kiel, el koordinata priskribo de fizika situacio per iu inercia referenca sistemo, oni trovas priskribon per alia inercia referenca sistemo. Tamen, tio ne montras tion kion observanto vere vidas. Por tio, oni devas konsideri plurajn faktorojn, ^cefe, Doppleran efikon. 

En anta^uaj artikoloj \cite{PaivaTeixeira2006,PaivaTeixeira2007a,PaivaTeixeira2007b} ni studis luman Doppleran efikon ^ce kelkaj specialaj situacioj: tie a^u observanto a^u fonto restas. ^Ci tie, amba^u movi^gas. Ili movi^gas ^ce konstanta propra akcelo, kiu tre ta^ugas por special-relativeco.  Fakte, tiu akcelo estas relativeca ekvivalenta de Newtona konstanta akcelo, kaj plibone ne generas rapidon pli granda ol rapido de lumo en vakuo, $c$; cetere ^gi multe simpligas kalkulon, kaj havas specialajn kvalitojn pri kiuj ni diskutos en Konkludo.  

Luma Dopplera efiko estas malsameco $\nu\,'\!\neq\!\nu$ de frekvenco $\nu\,'$ de el\-igita lumo kaj frekvenco $\nu$ de observata lumo, pro movado de fonto a^u observanto, a^u amba^u. Proporcio $D=\nu/\nu\,'$ nomi^gas Dopplera faktoro. ^Car frekvenco estas inverso de periodo, tial ni difinas Doppleran faktoron kiel
\bea                                                          \label{f01} 
D(\tau):=\frac{\dd\tau'}{\dd\tau} 
= \frac{\gamma(t)}{\gamma(t')}\frac{\dd t'}{\dd t} \ .
\eea 
^Ci tie $\dd\tau'$ estas infinitezima propra intertempo per fonto inter eligo de du lumaj signaloj, kaj $\dd\tau$ estas infinitezima propra intertempo per observanto inter enigo de tiuj signaloj. Momento de eligo de lumo-signalo el fonto nomi^gas $t'$ a^u $\tau'$, kaj momento de enigo de tiu signalo en observanto nomi^gas $t$ a^u $\tau$. ^Ce la tria termo en (\ref{f01}), ni uzis rilaton de intertempa dilato $\dd\tau=\dd t/\gamma$ inter propra kaj koordinata intertempoj, kie $\gamma:=1/\sqrt{1-v^2/c^2}$, kaj $v:=\dd x/\dd t$\ . 

Por kalkuli Doppleran faktoron ni unue kalkulas distancon inter loko de fonto je momento $t'$ de eligo de signalo kaj loko de observanto je momento $t$ de enigo de tiu signalo. Poste ni egalas tiun distancon al distanco $c(t-t')$ kiun lumo trakuras dum intertempo $t-t'$. Tiu ekvacio, nomota elig-eniga ekvacio, rilatas momentojn de eligo $t'$ (a^u $\tau'$) kun momentoj de enigo $t$ (a^u $\tau$). Diferenciante ^gin ni havigas Doppleran faktoron je momento $t$ (a^u $\tau$), per ekvacio~(\ref{f01}). 

Dopplera efiko kun $D < 1$ nomi^gas ``ru^g-deloki^go'', ^car frekvenco eti^gas; ekzemple, deloki^go de flava koloro al ru^ga. Kontra^ue, efiko kun $D>1$ nomi^gas ``viol-deloki^go''. Se $D=1$ oni diras ke ne estas Dopplera efiko. Rimarku, ke ^car propratempoj $\dd\tau'$ kaj $\dd\tau$ estas Lorentze skalaraj, do anka^u Dopplera faktoro $D$ estas Lorentze skalara.  

Iuj rezultoj pri Dopplera efiko trovi^gas ekzemple en~\cite{PaivaTeixeira2006,PaivaTeixeira2007a,PaivaTeixeira2007b,Rindler77}. Unue, se fonto kaj observanto kolinie fori^gas unu de la alio,  okazas ru^g-deloki^go. Kontra^ue, se ili kolinie alproksimi^gas, okazas viol-deloki^go. Due, se fonto ^cirka^uiras restantan observanton,  okazas ru^g-deloki^go, pro intertempa dilato ^ce fonto. Kontra^ue, se observanto ^cirka^uiras restantan fonton, okazas viol-deloki^go.   

\begin{figure}                                                   
\centerline{\epsfig{file=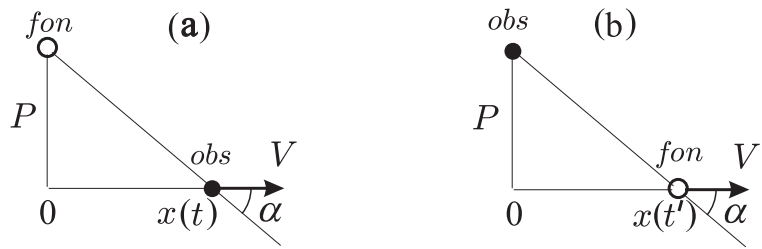,width=10cm}} 
\caption{El sekcio~\ref{e}. (a) Restanta fonto kaj movi^ganta observanto. 
(b) Kontra^ue. }
\label{Fig1} 
\end{figure}

Iom pli ^generale, esti\^gu restanta fonto kaj movi^ganta observanto kun rapido $V$, kiel en figuro~\ref{Fig1}(a), do elig-eniga ekvacio estas $\sqrt{P^2+x^2(t)} = c(t-t')$. Diferenciante kaj uzante iom da geometrio, Dopplera faktoro esti^gas 
\bea                                                          \label{f02} 
D=\gamma\left(1-\frac{V}{c}\cos\alpha\right)\ .  
\eea
Kontra^ue, por restanta observanto kaj movi^ganta fonto kun rapido $V$, kiel en figuro~\ref{Fig1}(b), elig-eniga ekvacio estas $\sqrt{P^2+x^2(t')} = c(t-t')$. Diferenciante, Dopplera faktoro esti^gas  
\bea                                                          \label{f03}
D=\left[\gamma\left(1+\frac{V}{c}\cos\alpha\right)\right]^{-1}\ .  
\eea
En tiuj du pli ^generalaj okazoj, Dopplera faktoro dependas de rapido kaj de angulo. 

Ni esploras kvin malsamajn fizikajn sistemojn, en kiuj observanto kaj fonto movi^gas kun sama konstanta propra akcelo. Je la unuaj du sistemoj, observanto kaj fonto movi^gas kolinie. Je la tria, movadoj estas paralelaj kaj ortaj al linio kunigante ilin. Kaj fine, je la du lastaj, movadoj estas anka^u paralelaj, sed klinaj rilate al tiu linio. ^Ce ^ciu sistemo, ni inspektas plurajn eblajn fazojn. ^Ce fazo~0, restanta observanto ricevas signalon eligitan el fonto anka^u restanta; ^car ne estas Dopplera efiko en tiu fazo ($D=1$), tiu fazo ne estos ordinare prezentata. ^Ce fazo~1, movi^ganta observanto ricevas signalon eligitan el ankora^u restanta fonto. Fine, ^ce fazo~2, movi^ganta observanto ricevas signalon eligitan el anka^u movi^ganta fonto; ni rimarkos, ke ^ci tiu fazo ne ^ciam ekzistas. 

Sekcio~\ref{kpa} difinas propran akcelon de korpo, kaj priskribas movadon de korpo se tiu akcelo estas konstanta. En sekcio~\ref{omf}, observanto movi^gas malanta^u fonto, kaj en sekcio~\ref{fmo}, kontra^ue. Sekcio~\ref{nm} priskribas la nekolinian ortan movadon. Sekcioj~\ref{omfn} kaj \ref{fmon} priskribas klinajn movadojn kun observanto malanta^u fonto kaj kontra^ue. Sekcio~\ref{k} konkludas.  

\section{Konstanta propra akcelo}                             \label{kpa}

Anta^uigante studon de Dopplera faktoro, ^ci tiu sekcio difinas propran akcelon, kaj priskribas movadon la^u konstanta propra akcelo, kiel M\o ller \cite[pa^goj 73-74]{Moller72} kaj Rindler \cite[pa^go 49]{Rindler77}.

Esti^gu inercia referenca sistemo en kiu korpo momente ripozas. En tiu sistemo, deriva^jo de rapido de korpo rilate al tempo  nomi^gas propra akcelo $a$ de korpo (rimarku, ke $a$ estas Lorentze skalara). ^Ce unudimensia movado, oni montras, ke en iu ajn inercia referenca sistemo 
\bea                                                          \label{f04}
a=\frac{\dd(\gamma v)}{\dd t}\ . 
\eea 

Esti^gu korpo komence restanta ($v_0=0$) en pozicio $0$ de akso $x$ de inercia sistemo $S_0$ de referenco. Ekde $t=0$ ($\tau=0$), ^gi movi^gu kun konstanta propra akcelo $a$ en pozitiva direkto de $x$. Integrante (\ref{f04}), la movado esti^gas   
\bea                                                          \label{f05}
x &=& \frac{c^2}{a}\left(\sqrt{1+(at/c)^2}-1\right)
=\frac{c^2}{a}\left[\,\cosh(a\tau/c)-1\,\right]\ ,
\\                                                            \label{f06}
v &=& \frac{at}{\sqrt{1+(at/c)^2}}=c\tanh(a\tau/c)\ , 
\\                                                            \label{f07}
\gamma &=& \sqrt{1+(at/c)^2}=\cosh(a\tau/c)\ ,
\\                                                            \label{f08} 
at/c &=& \sinh(a\tau/c)\ .
\eea 
Rimarku en (\ref{f06}), ke $v\rightarrow c$ kiam $t\rightarrow\infty$. Eblas pruvi, ke Newtona akcelo ($a_N:=\dd v/\dd t$) de tiu movado estas $a_N(t)=a/\gamma^3(t)$, kiu eti^gas la^u tempo, ^car $\gamma(t)$ egi^gas la^u rapido.  

\section{Observanto malanta^u fonto}                          \label{omf}

\begin{figure}                                                   
\centerline{\epsfig{file=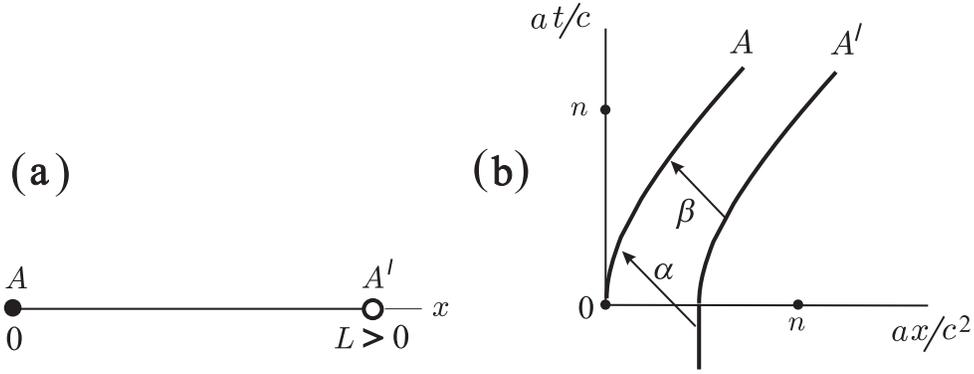,width=13cm}}
\caption{El sekcio \ref{omf}. (a) Observanto $A$ kaj luma fonto $A\,'$ komence restas ^ce $x=0$ kaj $x=L>0$, respektive. (b) Amba^u $A$ kaj $A'$ akceli^gas dekstren ekde $t=0$. Sagoj je 135$^{\rm o}$ indikas lumajn signalojn. Observanto ricevas signalojn eligitajn el fonto kaj restanta~($\alpha$) kaj movi^ganta~($\beta$). }
\label{Fig2} 
\end{figure}

La^u figuro~\ref{Fig2}(a), kaj observanto $A$ kaj luma fonto $A'$ restas  en inercia sistemo $S_0$ de referenco. Ekde $t=0$, amba^u movi^gas dekstren en akso $x$, kun sama konstanta propra akcelo $a$. 

\begin{figure}                                                   
\centerline{\epsfig{file=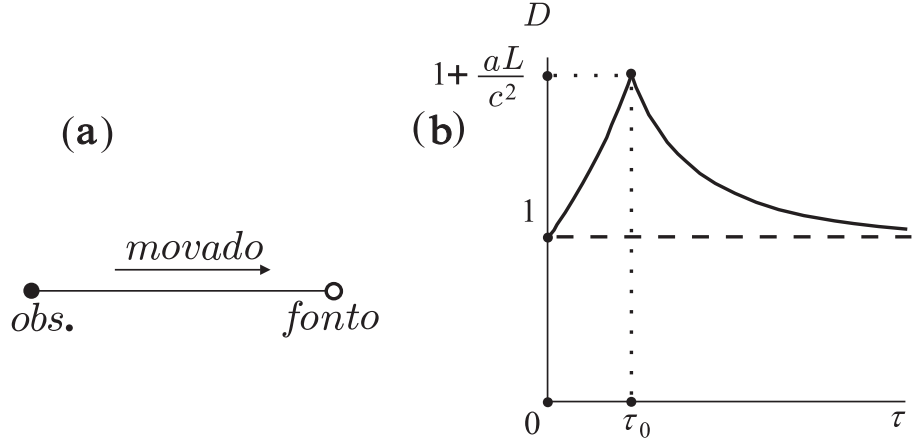,width=12cm}}
\caption{El sekcio~\ref{omf}. (a) Direkto de akcelo kaj movado. (b) Dopplera faktoro $D$ kiel funkcio de propratempo $\tau$ de observanto. Estas $\tau_0$ en~(\ref{f10}). }
\label{Fig3} 
\end{figure}

^Ce fazo~1, observanto movi^gas la^u (\ref{f05}), kaj ricevas signalojn (kiel sago $\alpha$ en figuro~\ref{Fig2}(b)) eligitajn el fonto restanta en $L>0$; do elig-eniga ekvacio estas
$L-x(t)=c(t-t')$. Uzante (\ref{f05}) por $x(t)$, kaj (\ref{f08}) por $t(\tau)$, kaj $t'=\tau'$, tiu ekvacio esti^gas
\bea                                                          \label{f09}
\tau'=\frac{c}{a}[\,{\rm exp}(a\tau/c)-(1+aL/c^2)\,]\ .
\eea 
Fazo~1 fini^gas je $\tau_0$, kiu estas la valoro de $\tau$ en (\ref{f09}) se $\tau'=0$:   
\bea                                                          \label{f10}
{\rm exp}\left(\frac{a\tau_0}{c}\right):=1+\frac{aL}{c^2}\ . 
\eea
Kalkulante $\dd\tau'/\dd\tau$ el (\ref{f09}) aperas Dopplera faktoro 
\bea                                                          \label{f11}
D_1={\rm exp}(a\tau/c), \hspace{2mm} 0<\tau<\tau_0\ .
\eea 
Oni rimarkas, ke $D$ eksponenciale egi^gas en fazo~1, kiel figuro~\ref{Fig3}(b) montras. Tio estas facile komprenebla, ^car en inercia sistemo $S_0$, observanto alproksimi^gas al restanta fonto pli kaj pli rapide, do li vidas viol-deloki^gon pli kaj pli forta. 

^Ce fazo~2 (kiel sago $\beta$ en figuro~\ref{Fig2}(b)), amba^u $A\,'$ kaj $A$ movi^gas, do nova elig-eniga ekvacio estas $L+x(t')-x(t)=c(t-t')$; uzante (\ref{f05}) por $x(t')$ kaj $x(t)$, kaj uzante (\ref{f08}) por $t$ kaj $t'$, tiu ekvacio esti^gas 
\bea                                                          \label{f12}
{\rm exp}(a\tau'/c)={\rm exp}(a\tau/c)-(aL/c^2).
\eea 
Fine, kalkulante $\dd\tau'/\dd\tau$ el (\ref{f12}) aperas faktoro  
\bea                                                          \label{f13}
D_2=\frac{1}{1-(aL/c^2)\,{\rm exp}(-a\tau/c)}\ , \hspace{2mm} \tau>\tau_0\ .
\eea

Ni rimarkas en (\ref{f13}) kaj en figuro~\ref{Fig3}(b), ke $D_2$ eti^gas ekde $\tau=\tau_0$. Tio estas facile komprenebla en inercia sistemo $S_0$ de referenco: Newtona akcelo $a_N(t)$ de observanto, je momento $t$ de ricevo de lumo, estas pli eta ol Newtona akcelo $a_N(t')$ de fonto, je momento $t'<t$ de eligo de tiu lumo; do  observanto en $x(t)$ alproksimi^gas al loko $x(t')$ de eligo de lumo pli kaj pli malrapide.   

Ni anka^u rimarkas en (\ref{f13}) kaj en figuro~\ref{Fig3}(b), ke okazas $D_2\rightarrow1$ kiam $\tau\rightarrow\infty$; tiam amba^u observanto kaj fonto havas rapidon ^cirka^ue $c$, kaj ne ekzistas Dopplera efiko. Newtona kinematiko facile eksplikas tiun fakton: se fonto kaj observanto estas kolinie movi^gantaj kun similaj rapidoj, Dopplera efiko ne ekzistas. Sed relativeca kinematiko bezonas fari pli detalan analizon. 
Rapido de observanto (je momento $\tau$ de ricevo de signalo) mezurata en inercia sistemo de momenta ripozo de fonto (je momento $\tau'$ de eligo de tiu signalo) estas $V=[v(\tau)-v(\tau')]/[1-v(\tau)v(\tau')/c^2]$. Uzante (\ref{f06}) por $v(\tau)$ kaj $v(\tau')$, tiu esti^gas $V=c\,\tanh\,a(\tau-\tau')/c$.  
Ekvacio (\ref{f12}) implicas $(\tau-\tau')\rightarrow (L/c)\,{\rm exp}(-a\tau/c)\rightarrow0$ kiam $\tau\rightarrow\infty$, do relativa rapido $V$ (de loko de ricevo al loko de eligo) nuli^gas kiam $\tau\rightarrow\infty$, eksplikante $D_2\rightarrow1$. 

Kurioze, ^ce inercia sistemo $S_0$ de referenco, distanco inter loko  $L+x(t')$ de eligo de signalo kaj loko $x(t)$ de ricevo ^ciam eti^gas, de $L$ je $t=0$ ^gis $L/2$ je $t\rightarrow\infty$. Tio estas facile komprenebla ^car rapido de observanto kiam $t\rightarrow\infty$ estas proksimume $c$, kaj rapido de signalo estas $-c$; do renkonto de observanto kun signalo okazas ^ce mezo de konstanta distanco $L$ de observanto al fonto. 

\section{Fonto malanta^u observanto}                          \label{fmo}

\begin{figure}                                                   
\centerline{\epsfig{file=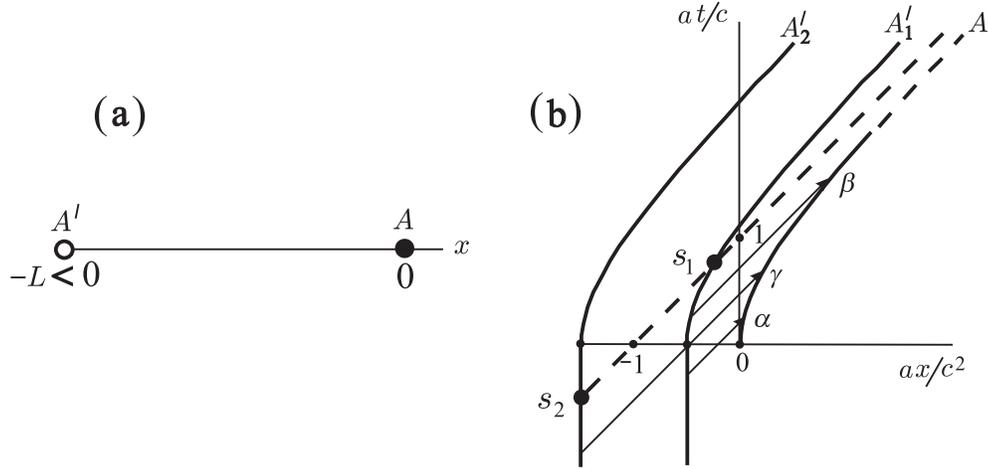,width=13cm}}
\caption{El sekcio~\ref{fmo}. (a) Fonto~$A\,'$ kaj observanto $A$ komence restas en $x=-L<0$ kaj $x=0$, respektive. (b) Amba^u akceli^gas dekstren ekde $t=0$; fontoj~$A_1'$ kaj~$A_2'$ movi^gas ekde $x=-L_1$ kaj $-L_2$, respektive, estante $0<L_1<c^2\!/a$ kaj $L_2>c^2\!/a$. Sagoj je~$45^{\rm o}$ indikas lumajn signalojn. Movi^ganta observanto~$A$ ricevas lumon eligitan el fonto~$A_1'$ kaj restanta~$(\alpha)$ kaj movi^ganta~$(\beta)$. Sed li ricevas lumon eligitan el fonto~$A_2'$ nur restanta~$(\gamma)$. Eventoj~$s_2$, a^u~$s_1$, estas lasta eligo de signalo atinganta~$A$. La strekata rekta linio estas asimptoto de hiperbola kurbo de observanto~$A$.}
\label{Fig4} 
\end{figure}

La^u figuro~\ref{Fig4}, observanto~$A$ komence restas  en $x=0$, kaj fonto $A_1'$ a^u $A_2'$ komence restas en $-L_1$ a^u $-L_2$, respektive, estante  $0<L_1<c^2/a$ kaj $L_2>c^2/a$. Por fazo~1, elig-eniga ekvacio estas $L+x(t)=c(t-t')$. Tiu fazo da^uras ^gis momento $\tau_0$ la^u
\bea                                                          \label{f14}
{\rm exp}\left(-\frac{a\tau_0}{c}\right):=1-aL/c^2 \hspace{3mm} {\rm se}\hspace{2mm} aL/c^2<1. 
\eea 
Rimarku, ke se $L>c^2/a$ (kiel por fonto $A_2'$ ^ce figuro~\ref{Fig4}(b)), fazo~1 da^uras ^gis $\tau\rightarrow\infty$, kaj do fazo~2 ne ekzistas. En tia okazo, lasta signalo $s_2$ en figuro~\ref{Fig4}(b), kiu atingas observanton $A$ (je $\tau\rightarrow\infty$) estas eligita je $t'_2=-(1/c)(L-c^2/a)$. Signaloj eligitaj post $t'_2$ ne atingas observanton kiun ekmovi^gis sufi^ce fore de fonto ($L>c^2/a$). Se tamen $0<L<c^2/a$, fazo~2 ekzistas. 

\begin{figure}                                                   
\epsfig{file=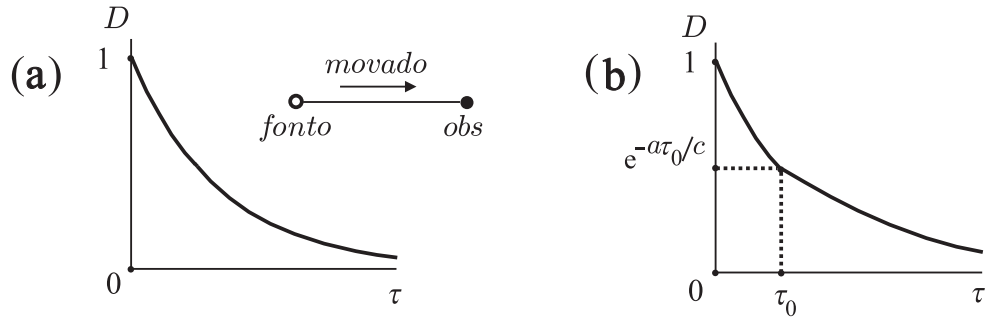,width=13cm}
\caption{El sekcio~\ref{fmo}. Dopplera faktoro $D$ kiel funkcio de propra tempo $\tau$ de observanto. (a) Okazo $L>c^2\!/a$ (kiel por fonto $A_2'$ en figuro~\ref{Fig4}(b)), havanta nur unu fazon. (b) Okazo $0<L<c^2\!/a$ (kiel por fonto $A_1'$ en figuro~\ref{Fig4}(b)), havanta du fazojn. Estas $\tau_0$ en~(\ref{f14}). }
\label{Fig5} 
\end{figure}

Por fazo~1, el elig-eniga ekvacio, Dopplera faktoro $\dd\tau'/\dd\tau$ esti^gas (vidu figuron~\ref{Fig5})    
\bea                                                          \label{f15}
D_1={\rm exp}(-\frac{a\tau}{c}), \hspace{2mm} 0<\tau<\tau_0\ ,
\eea 
$\tau_0$ estante momento de ricevo de lumo eligita je $\tau'=0$. Estas facile komprenebla, en inercia sistemo $S_0$ de referenco, kial $D_1(\tau)$ eksponenciale eti^gas: observanto iras pli kaj pli rapide fore de  restanta loko de eligo de lumo. 

Por fazo~2, elig-eniga ekvacio estas $L+x(t)-x(t')=c(t-t')$, do Dopplera faktoro $\dd\tau'/\dd\tau$ esti^gas (vidu figuron~\ref{Fig5}(b))    
\bea                                                          \label{f16}
D_2=\frac{1}{1+(aL/c^2)\exp(a\tau/c)}\ , 
\hspace{2mm} 0<L<c^2/a\ , \hspace{2mm} \tau>\tau_0\ . 
\eea
En tia okazo, signalo eligita je $\tau'_1:=(c/a)\,{\rm ln}(c^2/aL)$ estas la lasta kiu sukcesas atingi observanton (kiam $\tau\rightarrow\infty$). Vidu $s_1$ en figuro~\ref{Fig4}(b).

Ni rimarkas en figuro~\ref{Fig5}(b), ke $D_2\rightarrow0$ kiam $\tau\rightarrow\infty$; tiam rapido de observanto ^cirka^uas~$c$, kaj li ricevas signalon eligitan je momento $\tau\,'_1$. Je ^ci tiu finhava tempo, rapido de fonto estas anka^u pozitiva, sed ne ^cirka^uas~$c$\,; tio klarigas ekstreman ru^g-deloki^gon $D_2\rightarrow0$. 

Oni montras, ke je $\tau=\tau_0$, okasas $\dd D_1/\dd\tau=-(a/c)(1-aL/c^2)$ kaj $\dd D_2/\dd\tau=-(a/c)(1-aL/c^2)aL/c^2$. ^Car $aL/c^2<1$, je $\tau_0$, grafika^jo estas malpli klinata ^ce fazo~2 ol ^ce fazo~1, kiel montras figuro~\ref{Fig5}(b). Tio okazas ^car ekde $\tau_0$, ricevataj signaloj estas eligitaj el jam movi^ganta fonto.    

\section{Nekoliniaj movadoj}                                   \label{nm} 

\begin{figure}                                                   
\centerline{\epsfig{file=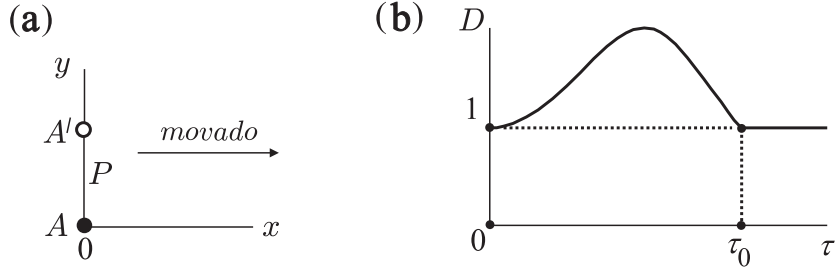,width=11cm}} 
\caption{El sekcio~\ref{nm}. (a) Komenca pozicio de fonto $A\,'$ kaj observanto $A$. (b) Dopplera faktoro $D$ kiel funkcio de propratempo $\tau$ de observanto. Estas  $\tau_0$ en~(\ref{f18}).}
\label{Fig6} 
\end{figure}

La^u figuro~\ref{Fig6}(a), observanto $A$ restas en pozicio $[x,y]=[0,0]$ kaj luma fonto $A\,'$ restas en pozicio $[0,P]$. Ekde $t=0$, amba^u movi^gas dekstren kun sama konstanta propra akcelo~$a$. Komence,  movi^ganta observanto ricevas lumon el ankora^u restanta fonto, do elig-eniga ekvacio estas $\sqrt{P^2+x^2(t)}=c(t-t')$. Uzante (\ref{f05}) por $x(t)$, (\ref{f08}) por $t(\tau)$, kaj $t'=\tau'$, poste kalkulante $\dd\tau'/\dd\tau$, aperas Dopplera faktoro ^ce fazo~1 (vidu figuron~\ref{Fig6}(b))   
\bea                                                          \label{f17}
D_1=\cosh(a\tau/c)\left(1-\frac{\tanh(a\tau/c)(\cosh(a\tau/c)-1)}{\sqrt{(aP/c^2)^2+(\cosh(a\tau/c)-1)^2}}\right)\ , \hspace{3mm} 0<\tau<\tau_0\ , 
\eea
$\tau_0$ estante momento de ricevo de lumo eligita je $\tau'=0$: 
\bea                                                          \label{f18}
\cosh(a\tau_0/c):=1+\frac{1}{2}(aP/c^2)^2\ .
\eea 

Ni analizas tiun grafika^jon ^cirka^u $\tau=0$. Tiuokaze, el (\ref{f17}), okazas $D_1\approx1+(1/2)(a\tau/c)^2$. Ni rimarkas, ke je $\tau=0$, okazas $\dd D_1/\dd\tau=0$ kaj $\dd^2 D_1/\dd\tau^2=(a/c)^2$. Do tiu grafika^jo komencas horizontale kaj kurbi^gas supren (^gia kurbeco ne dependas de $P$). La interpreto de tiu rezulto estas facila. Komence okazas orta Dopplera efiko, kiu ne ekzistas en Newtona kinematiko. Tiu ordinare faras etan viol-deloki^gon, kiu egi^gas ^car rapido egi^gas. Poste, la for-movado aperigas faktoron de ru^g-deloki^go. Fine ^ci tiuj du malaj efektoj nuligas unu la alion je $\tau=\tau_0$, ^guste kiam lasta signalo de fonto ankora^u ripozanta atingas observanton. 
 
^Ce fazo~2, amba^u observanto kaj fonto estas movi^gantaj, do elig-eniga ekvacio estas \linebreak[4] $\sqrt{P^2+[x(t)-x(t')]^2}=c(t-t')$. Uzante (\ref{f05}) por $x(t)$ kaj $x(t')$, kaj uzante (\ref{f08}) por $t$ kaj $t'$, tiu ekvacio multe simpli^gas al    
\bea                                                          \label{f19}
\tau'=\tau-\tau_0\ ; 
\eea 
do $D_2=1$, t.e., ne ekzistas Dopplera efiko ^ce fazo~2. Vidu figuron~\ref{Fig6}(b). 
 
Nun ni diskutas ^ci tiun konstantecon de Dopplera faktoro. Kiel ni raportis en Enkonduko, se observanto pasas orte restantan fonton, okazas viol-deloki^go pro tempa dilato je rapido de observanto. Alie, se observanto iras kolinie fore de restanta fonto, okazas ru^g-deloki^go. En ^ci tiu sekcio, tiuj du komponoj konkursas. Tiamaniere Dopplera efiko dependas de rapido de observanto rilate al fonto, kaj de angulo inter tiu rapido kaj direkto de signal-eligo, kiel en figuro~\ref{Fig1}(a). Konsideru unue la rapidon. Same kiel en sekcio~\ref{omf}, la relativa rapido estas $V=c\tanh a(\tau-\tau')/c$. Pro ekvacio~(\ref{f19}), $V$ estas konstanta. Due, konsideru la angulon. En inercia referenca sistemo $S_0$, okazas $\sin\alpha=P/[c(t-t')]$. La^u Lorentza transformo, en inercia sistemo de momenta ripozo de fonto, la diferenco de tempoj esti^gas
\bea                                                          \label{f20}
\gamma(t')\left((t-t')-\frac{v(t')}{c^2}[x(t)-x(t')]\right)
= \frac{c}{a}\sinh \frac{a(\tau-\tau')}{c}
\eea
kie ni uzis (3)--(6) por trovi la dekstran termon. Pro ekvacio~(\ref{f19}) denove, $\alpha$ havas la saman valoron en ^ciu inercia sistemo de momenta ripozo de fonto, eksplikante kial la Dopplera faktoro estas konstanta.

Ni ^jus diskutis pri la konstanteco de Dopplera faktoro en fazo~2. Pri ^gia valoro, notu ke $D(\tau_0)=1$ por iu ajn valoro de $P$. Tiun interesan (kaj misteran?) fakton prezentas plidetale nia estonta artikolo {\em Dopplera efiko de luma ebeno vidata per akcelata korpo}. ^Ci tie en Konkludo, ni rediskutis pri konstanteco kaj nuleco de $D_2$ uzante he^uristikajn argumentojn.

\section{Observanto malanta^u fonto, nekolinie}              \label{omfn}

\begin{figure}                                                   
\centerline{\epsfig{file=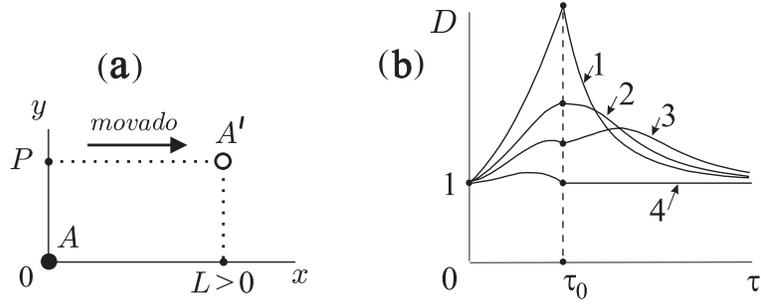,width=10cm}} 
\caption{El sekcio~\ref{omfn}. (a) Komencaj pozicioj de observanto $A$ kaj fonto $A\,'$. 
(b) Ekzemploj de Dopplera faktoro kiel funkcio de propratempo de observanto, por $\cosh a\tau_0/c=2$. Valoro de $[\,aL/c^2,aP/c^2\,]$ en ^ciu ekzemplo estas: {\bf 1}: $[1+\sqrt{3}/2,3/2]$; {\bf 2}: $[1,\sqrt{3}]$; {\bf 3}: $[1/2,\sqrt{11}/2]$; {\bf 4}: $[0,\sqrt{2}]$. }
\label{Fig7} 
\end{figure}

Ni ^generaligas sekciojn~\ref{omf} kaj \ref{nm}. Vidu en figuro~\ref{Fig7}(a), ke observanto $A$ kaj fonto $A\,'$ komence restas  en $[x,y]=[0,0]$ kaj $[L,P]$, respektive, $L$ estante pozitiva kaj $P$ ioma. Ekde $t=0$, amba^u movi^gas dekstren kun sama konstanta propra akcelo $a$. Por fazo~1, elig-eniga ekvacio estas $\sqrt{P^2+[L-x(t)]^2}=c(t-t')$. Tiu fazo da^uras ^gis momento $\tau_0$ la^u  
\bea                                                          \label{f21} 
\cosh\frac{a\tau_0}{c}:=1+\frac{a^2(L^2+P^2)/c^4}{2(1+aL/c^2)}\ . 
\eea
Interesas scii kie estas observanto kiam fazo~1 fini^gas. Tio dependas de valoroj de $P$, kaj $L$, kaj $a$. Fakte, uzante (\ref{f05}) kaj (\ref{f21}), oni pruvas ke la signumo de $x(\tau_0)-L$ estas 
\bea                                                          \label{f22}
\sigma:={\rm sign}(P^2-L^2-2c^2L/a)\ .  
\eea
Do fazo~1 fini^gas kiam observanto estas en a^u $x(\tau_0)<L$ a^u $x(\tau_0)=L$ a^u $x(\tau_0)>L$ la^u $\sigma$ estas malpozitiva a^u nula a^u pozitiva, respektive. Tio estas facile komprenebla, ^car des pli granda $|P|$, des pli malfrua estas ricevo de lumo, do des pli granda estas $x(\tau_0)$.

Por fazo~1, el la supra elig-eniga ekvacio, Dopplera faktoro esti^gas  
\bea                                                          \label{f23}
D_1=\cosh(a\tau/c)\left(1-\frac{\alpha_1\,\tanh( a\tau/c)}{\sqrt{{\alpha_1}^2+(aP/c^2)^2}}\right)\ , \hspace{3mm} 0<\tau<\tau_0\ , 
\eea 
estante
\bea                                                          \label{f24}
\alpha_1:=\cosh a\tau/c-1-aL/c^2\ . 
\eea 
Vidu figuron~\ref{Fig7}(b). Eblas pruvi, ke $\dd D_1/\dd\tau=(aL/c)/\sqrt{L^2+P^2}$ je $\tau=0$; t.e., $\dd D_1/\dd\tau$  estas pozitiva (krom se $L=0$ kiel ^ce kurbo ${\bf 4}$, kiu ripetas figuron~\ref{Fig6}(b) el sekcio~\ref{nm}). La komence kreskanta valoro de $D_1(\tau)$ okazas pro la eganta rapido de alproksimi^gado de observanto al fonto. Eblas pruvi anka^u, ke denove  $\dd^2D_1/\dd\tau^2=(a/c)^2$ je $\tau=0$, same kiel en sistemo de sekcio~\ref{nm}; do la komenca kurbeco de $D_1(\tau)$ estas supren, kaj dependas nek de $L$ nek de $P$. 

Ni vidas en figuro~\ref{Fig7}(b), ke la klino de $D_1(\tau)$, je $\tau_0$, estas a^u pozitiva (kurbo~{\bf 1}), a^u nula~({\bf 2}), a^u malpozitiva~({\bf 3}). Simpla kalkulo pruvas, ke tiuj signumoj estas $-\sigma$. La ^generalaj formoj de tiuj grafika^joj estas kombinoj de grafika^joj de
figuro~\ref{Fig3}(b) kun figuro~\ref{Fig6}(b). Fakte, se $L$ estas sufi^ce granda, figuro~\ref{Fig3}(b) superas, kaj kontra^ue se $P$ estas sufi^ce granda, figuro~\ref{Fig6}(b) superas.  

^Ce fazo~2, esti^gas $\sqrt{P^2+[L+x(t')-x(t)]^2}=c(t-t')$; uzu (\ref{f05}) por $x(\tau)$ kaj $x(\tau')$, kaj uzu (\ref{f08}) por $t(\tau)$ kaj $t'(\tau')$, poste kalkulu Doppleran faktoron $\dd\tau'/\dd\tau$: 
\bea                                                          \label{f25}
D_2=\frac{\sinh(a\tau/c-a\tau'/c)+(aL/c^2)\sinh a\tau/c} {\sinh(a\tau/c-a\tau'/c)+(aL/c^2)\sinh a\tau'/c}\ , 
\hspace{2mm} \tau_0<\tau<\infty\ , 
\eea 
estante
\bea                                                          \label{f26}
\cosh a\tau'/c &=& \frac{1}{\alpha_2}\left(\alpha_3(1+\alpha_1)+\sqrt{{\alpha_3}^2+\alpha_2}\sinh a\tau/c\right)\ ,  
\\                                                            \label{f27}
\alpha_2 &:=& 2(aL/c^2)\cosh a\tau/c-1-(aL/c^2)^2\ , 
\\                                                            \label{f28} 
\alpha_3 &:=& \frac{1}{2}\left(\alpha_2-1-(aP/c^2)^2\right)\ ; 
\eea
vidu figuron~\ref{Fig7}(b). Same kiel en figuroj~\ref{Fig3}(b) kaj~\ref{Fig6}(b), ^ci tie $D_2(\infty)=1$. Figuro~\ref{Fig7}(b) anka^u montras, ke klino de $D_2(\tau)$, je $\tau_0$, estas a^u malpozitiva, a^u nula, a^u pozitiva, kaj eblas pruvi, ke tiuj signumoj estas $\sigma$. Fakte, se fazo~2 komencas kiam observanto estas en $x(\tau_0)<L$, je tiu momento fonto foriras kun Newtona akcelo pligranda ol tiu de observanto. Do la viol-deloki^go malforti^gas. Kontra^ue, se observanto estas en $x(\tau_0)>L$, je tiu momento, fonto alproksimi^gas al li, do la viol-deloki^go forti^gas.     

\section{Fonto malanta^u observanto, nekolinie}              \label{fmon}

\begin{figure}                                                   
\centerline{\epsfig{file=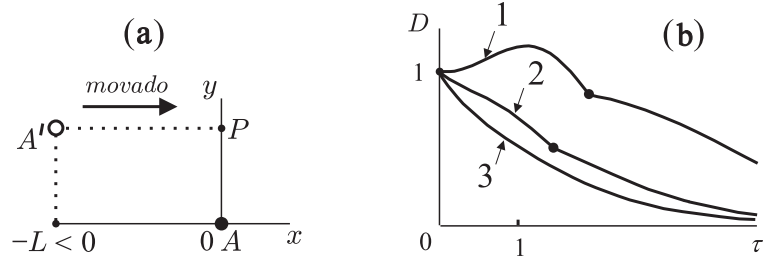,width=10cm}} 
\caption{El sekcio~\ref{fmon}. (a) Komenca pozicio de fonto $A\,'$ kaj de observanto $A$. 
(b) Ekzemploj de Dopplera faktoro kiel funkcio de propratempo de observanto. Valoroj de paro $[\,aL/c^2,aP/c^2\,]$ en ^ciu ekzemplo estas: {\bf 1}: $[\,3/20,2\,]$; {\bf 2}: $[\,1/2,1\,]$; {\bf 3}: $[\,5/4,1\,]$. En ekzemplo {\bf 3}, $aL/c^2>1$, do ^ci tiu okazo ne prezentas fazon~2. }
\label{Fig8} 
\end{figure} 

^Ci tiu sekcio ^generaligas sekciojn~\ref{fmo} kaj \ref{nm}. Vidu en figuro~\ref{Fig8}(a), ke observanto $A$ kaj luma fonto $A'$ komence restas respektive en $[x,y]=[0,0]$ kaj $[\,-L,P\,]$, $L$ estante pozitiva kaj $P$ ioma. Ekde $t=0$, amba^u movi^gas dekstren je sama konstanta propra akcelo $a$. Elig-enigaj ekvacioj de fazo~1 kaj~2 estas tiuj de sekcio~\ref{omfn} per anstata^uigo de $L$ al $-L$, do ekvacioj (\ref{f23})--(\ref{f28}) pravas anka^u ^ci tie, per tiu anstata^uigo.   

Fazo~1 da^uras ^gis momento $\tau_0$, la^u ekvacio~(\ref{f21}) (kun $-L$ anstata^u $L$). Same kiel en sekcio~\ref{omfn}, ni vidas, ke bezonas distingi du eblecojn. Se $L>c^2/a$, fazo~1 etendi^gas ^gis  $\tau\rightarrow\infty$, kiel en kurbo~{\bf 3} de figuro~\ref{Fig8}(b); en tia okazo, lumo eligita nur anta^u $t'_2=-(1/c)(L-c^2/a)$ atingas observanton, same kiel en sekcio~\ref{fmo}. Se tamen $0<L<c^2\!/a$, fazo~2 ekzistas. Tio estas videbla en kurboj~{\bf 1} kaj~{\bf 2} de figuro~\ref{Fig8}(b).    

Dum fazo~1, Dopplera faktoro estas la^u ekvacio~(\ref{f23}) (kun $-L$ anstata^u $L$). Vidu en figuro~\ref{Fig8}(b), ke ^ciuj kurboj komencas malsupren; tio okazas ^car observanto komence iras kun eganta rapido, fore de loko de lum-eligo. Fakte, simpla kalkulo pruvas, ke $\dd D_1/\dd\tau= -(aL/c)/\sqrt{L^2+P^2}$ kaj $\dd^2D_1/\dd\tau^2=(a/c)^2$ je $\tau=0$. 

Dum fazo~2, Dopplera faktoro estas la^u ekvacio~(\ref{f25}) (kun $-L$ anstata^u $L$). El tiu ekvacio, $D(\tau_0)=1-aL/c^2<1$, do je $\tau_0$, Dopplera efiko estas ru^g-deloki^go. 

Same kiel en sekcio~{\ref{fmo}, ekzistas signaloj kiuj neniam atingas  observanton. Fakte, la lasta signalo atinganta observanton (je $\tau\rightarrow\infty$) estas eligita je $\tau'_1=(c/a)\,{\rm ln}(c^2/aL)$. Do observanto kun rapido ^cirka^u $c$ ricevas signalon eligata el fonto kun eta rapido, eksplikante $D_2\rightarrow0$. Fine rimarku, ke ^ciu grafika^jo en figuro~\ref{Fig8}(b) estas kombino de figuro~\ref{Fig5}(b) kun figuro~\ref{Fig6}(b); speciale, $D_2(\infty)=0$.
 
\section{Konkludo}                                              \label{k}

^Ci tiu artikolo zorgis pri Doppleran efikon se luma fonto kaj observanto movi^gadas paralele, amba^u kun sama konstanta propra akcelo, kaj komencante el restado je sama momento. Ni vidis, ke se observanto movi^gas malanta^u fonto (sekcioj~\ref{omf} kaj \ref{omfn}), komence okazas viol-deloki^go, kiu i^gas nula efiko kiam $t\rightarrow\infty$ (figuroj~\ref{Fig3}(b) kaj~\ref{Fig7}(b)). Se, kontra^ue, fonto movi^gas malanta^u observanto (sekcioj~\ref{fmo} kaj \ref{fmon}), komence okazas  finia ru^g-deloki^go, kiu i^gas nefinia ru^g-deloki^go ($D\rightarrow0$) kiam $t\rightarrow\infty$ (figuroj~\ref{Fig5} kaj~\ref{Fig8}(b)). Rimarku, ke, en sekcio~\ref{fmon}, povas ekzisti viol-deloki^go inter la du ru^g-deloki^goj. Speciale, se observanto kaj fonto movi^gas orte al linio kunigante ilin (sekcio~\ref{nm}), figuro~\ref{Fig6}(b) montris, ke komence estas viol-deloki^go, kiu i^gas nula efiko ekde $t=t_0$.

Nun, ni klarigas konstantecon kaj neeston ($D=1$) de Dopplera efiko ^ce fazo~2 de orta-nekolinia movado de sekcio~\ref{nm}.
Per inercia sistemo de momenta ripozo de fonto kaj observanto, restanta fonto en $[x,y]=[0,P]$, kun propra akcelo $a$ la^u $x$, eligas lum-signalon. 
Je tiu momento, restanta observanto en $[x,y]=[0,0]$ havas propran akcelon $a$ la^u $x$. 
La valoro de $\tau-\tau'$ dependas de ^ci tiuj supraj kondi^coj. 
Iom poste, per nova inercia sistemo de momenta ripozo de fonto kaj observanto, propraj akceloj, rapidoj (nulaj) kaj relativaj pozicioj estas samaj.
^Car la kondi^coj estas samaj, do la valoro de $\tau-\tau'$ estas la sama, kaj do $D=1$.
Rimarku, ke ^ci tiu argumento validas nur:
(i) se propraj akceloj estas konstantaj (memoru, ke la^u sekcio~\ref{kpa}, propra akcelo estas difinata per inercia sistemo de momenta ripozo),
(ii) se movado estas orta, ^car kontra^ue, malsamaj Lorentzaj eti^gadoj de distanco $L$ aperas,
(iii) ^ce fazo~2, ^car kontra^ue, la kondi^coj ne estas samaj per sekvantaj inerciaj sistemoj.

Estas mirinda, ta^ugeco de konstanta propra akcelo al Dopplera efiko, ^ce speciala relativeco; tiu ta^ugeco estos plidetale vidata en estonta artikolo. 

Lasta komento estas: ni balda^u publikigos parton {\em II} de ^ci tiu artikolo, rilate al luma Dopplera efiko inter korpoj paralele movi^gantaj, amba^u je (malsamaj) konstantaj propraj akceloj; speciale ni traktas pri rigida movado~\cite{PaivaTeixeira2006}.

\subsection*{Gratuloj}

Ni kore dankas al Leo Squallyorbit pro sugesti studon de Dopplera efiko de observanto kaj fonto amba^u akcelataj.

\end{document}